\documentclass[a4paper]{jpconf}
\usepackage{graphicx}
\usepackage{amssymb,amsbsy,amsmath}
\begin{document}
%\title{Thermodynamics of frustrated $J_1$-$J_2$ quantum Heisenberg
%magnets: High-temperature expansion
%revisited
%}
\title{Magnetic susceptibility of frustrated spin-$s$ $J_1$-$J_2$ quantum Heisenberg
magnets: High-temperature expansion and exact
diagonalization data
}

\author{J. Richter$^1$, A. Lohmann$^1$, H.-J. Schmidt$^2$, D.C. Johnston$^3$}

\address{
$^1$Institut f\"ur Theoretische Physik, Otto-von-Guericke-Universit\"at
Magdeburg,\\
PF 4120, D - 39016 Magdeburg, Germany\\
$^2$Universit\"at Osnabr\"uck, Fachbereich Physik,
Barbarastr. 7, D - 49069 Osnabr\"uck, Germany\\
$^3$
Ames Laboratory and Department of Physics and Astronomy, Iowa State
University, Ames, Iowa 50011}

\ead{johannes.richter@physik.uni-magdeburg.de}

\begin{abstract}
   Motivated by recent experiments on low-dimensional frustrated quantum
 magnets with competing nearest-neighbor exchange coupling $J_1$ and next
 nearest-neighbor exchange coupling $J_2$ we investigate the magnetic
 susceptibility of two-dimensional $J_1$-$J_2$ Heisenberg models with arbitrary
 spin quantum number $s$. We use exact diagonalization and high-temperature
 expansion up to order 10 to analyze the influence of the frustration
 strength $J_2$/$J_1$ and the spin quantum number $s$ on the position and the height
 of the maximum of the susceptibility. The derived theoretical data can be
 used to get information on the ratio $J_2$/$J_1$ by comparing with susceptibility
 measurements on corresponding magnetic compounds.
\end{abstract}

\section{Introduction}
The investigation of frustrated magnetic systems is currently a field of
active theoretical and experimental research \cite{1,1a}. Systems with competing
nearest-neighbor (NN) exchange coupling $J_1$ and next nearest-neighbor (NNN)
exchange coupling $J_2$ can serve as model systems to study the interplay of
quantum effects, thermal fluctuations and frustration. The quantum $J_1$-$J_2$
Heisenberg models on the
square-lattice
exhibit several ground-state phases including non-classical
non-magnetic ground states, see, e.g., \cite{2}.
The corresponding  Hamiltonian reads
\begin{equation}\label{hamiltonian}
  H=J_1\sum_{\langle i,j\rangle} {\bf S}_i \cdot {\bf S}_j+J_2\sum_{[i,j]} {\bf
 S}_i\cdot {\bf S}_j ,
\end{equation}
where $({\bf S}_i)^2=s(s+1)$,  and $\langle i,j\rangle$ denotes NN
and $[i,j]$ denotes NNN bonds. For antiferromagnetic NNN bonds,
$J_2>0$, the spin system is frustrated irrespective of the sign of
$J_1$. Due to frustration the theoretical treatment
of this model is challenging.

The
numerous theoretical studies of  the ground state phase diagram so far did
not lead to a  consensus on the nature of the quantum ground state and on
the nature of the quantum phase transitions present in the model, see, e.g.,
\cite{3} and references therein. Interestingly there are also various compounds well described by
square-lattice $J_1$-$J_2$ Heisenberg models, such as oxovanadates \cite{4} and
iron pnictides \cite{5}. In experiments, typically temperature-dependent quantities are
reported. Hence reliable (and  flexible) tools are desirable to calculate
thermodynamic quantities such as the uniform magnetic susceptibility $\chi$.
In this paper we present two methods, namely the full exact diagonalization,
see, e.g., \cite{6},  and the high-temperature expansion \cite{7,8,9} to calculate
the temperature dependence of the magnetic susceptibility for the
square-lattice $J_1$-$J_2$ spin-$s$ Heisenberg model with both ferromagnetic (FM)
and antiferromagnetic (AFM) NN coupling $J_1$ and AFM NNN bonds $J_2$ for
arbitrary spin quantum number $s$. In particular, we analyze the position and
the height of the maximum in the susceptibility in dependence on $J_1$, $J_2$ and
$s$.

\begin{figure}[ht]
\includegraphics[width=37pc]{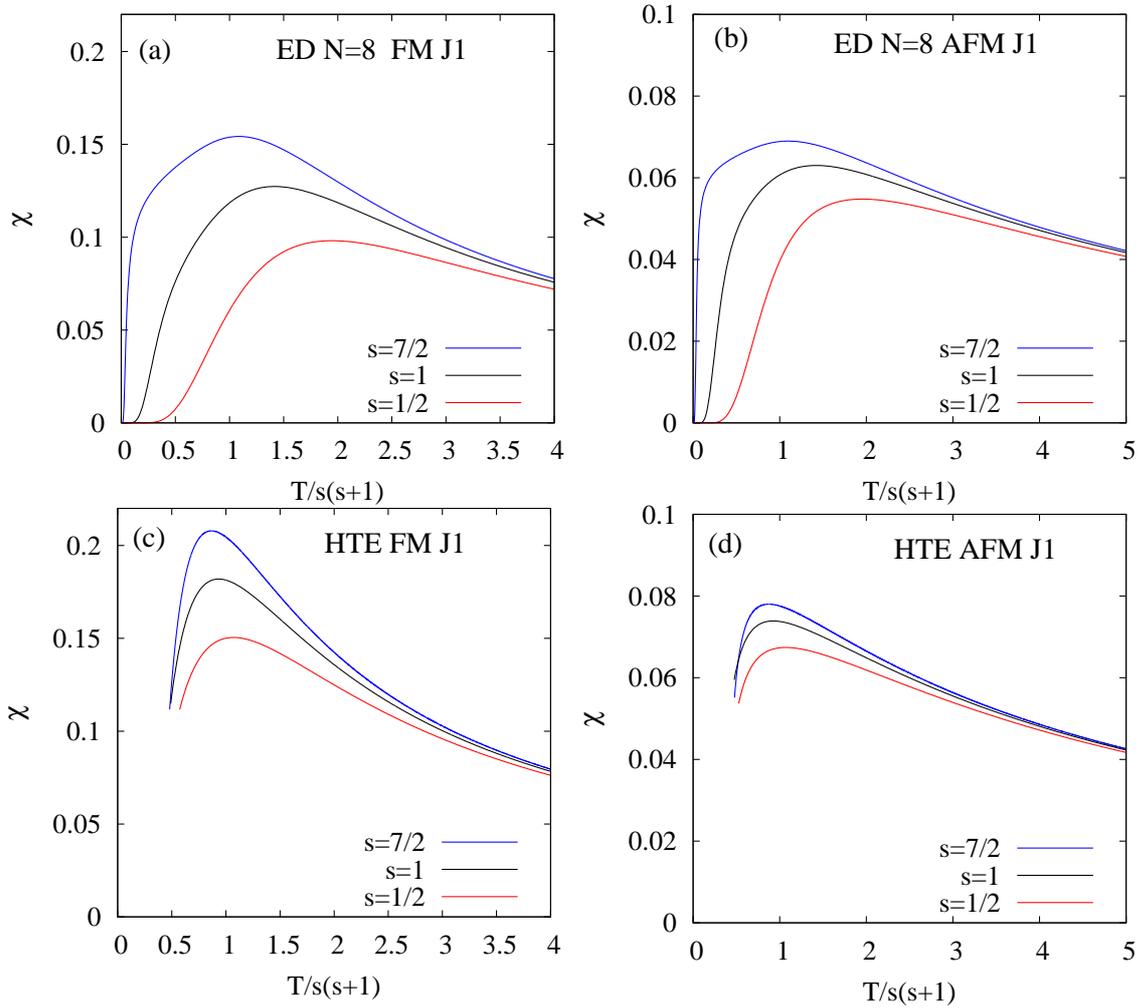}
\caption{\label{fig1}
Uniform susceptibility $\chi$ as a function of renormalized temperature
$T/s(s+1)$ for NNN exchange $J_2=1$ and three values of the spin quantum
number $s=1/2, 1$, and $7/2$.
(a) Numerical exact data for a finite square
lattice of $N=8$ sites and  antiferromagnetic $J_1=1$.
(b) Numerical exact data for a finite square
lattice of $N=8$ sites and  ferromagnetic $J_1=-1$.
(c) [6,4] Pad\'e approximant of
the 10th order HTE series for  an infinite square
lattice and  ferromagnetic $J_1=-1$.
(d) [6,4] Pad\'e approximant of
the 10th order HTE series for  an infinite square
lattice and  antiferromagnetic $J_1=1$.
The order of labeling in each  legend is the same as the order of the plots
(top to bottom).
}
\end{figure}

\section{Methods}

The full exact diagonaliazion (ED) yields numerical exact results at
arbitrary temperature $T$, but it is typically limited to about
$N=22$ sites for $s=1/2$ models. For larger spin quantum numbers $s$
the system size  $N$ accessible for  ED shrinks significantly.
Hence, ED is used preferably for $s=1/2$ and $s=1$. In the present
study we exploit the special symmetry properties  of the finite
square-lattice of $N=8$ sites and perform full ED for the
$J_1$-$J_2$ model for $s=1/2,1,..,9/2$, thus allowing to study the
role of the spin quantum number. Since the ED approach suffers from
the finite-size effect, the ED calculations do not yield
quantitatively correct results for the thermodynamic limit.
Nevertheless, they will give insight into the qualitative behavior
of the susceptibility. The high-temperature expansion (HTE) for the
$J_1$-$J_2$ model up to 10th order was presented in \cite{8},
however, restricted to $s=1/2$. This restriction can be overcome by
using our general HTE scheme for Heisenberg models with arbitrary
exchange patterns and arbitrary spin quantum number $s$ up to order
8 \cite{9}. The scheme is encoded in a simple C++-program and can be
downloaded \cite{10} and  freely used by interested researchers.

Very recently the present authors have extended this general HTE
scheme up to 10th order \cite{11}. Here we use this 10th order HTE
as an alternative method  to the ED.  We use here three different
subsequent Pad\'e approximants, namely Pad\'e [4,6], [5,5], and
[6,4], see e.g. \cite{7,9}. Such a Pad\'e approximant extends the
region of validity of the HTE series down to lower temperatures.
Since the HTE approach is designed for infinite systems  the HTE
data for the susceptibility maximum, in principle, can be
quantitatively correct, if the maximum is not located at too low
temperatures. Indeed, it was found \cite{9} that for the
unfrustrated  ($J_2=0$) square-lattice spin-$1/2$ Heisenberg
antiferromagnet the Pad\'e [4,4] approximant of the 8th order HTE series yields correct data for
the susceptibility maximum located at $T \approx 0.94 J_1$. However,
it may happen that a certain Pad\'e approximant does not work for
some particular values of $J_1$, $J_2$, and $s$, since Pad\'e
approximants may exhibit unphysical poles  for temperatures in the
region of interest. Hence we show in the next section only those
Pad\'e data not influenced by poles.

\begin{figure}[ht]
\includegraphics[width=40pc]{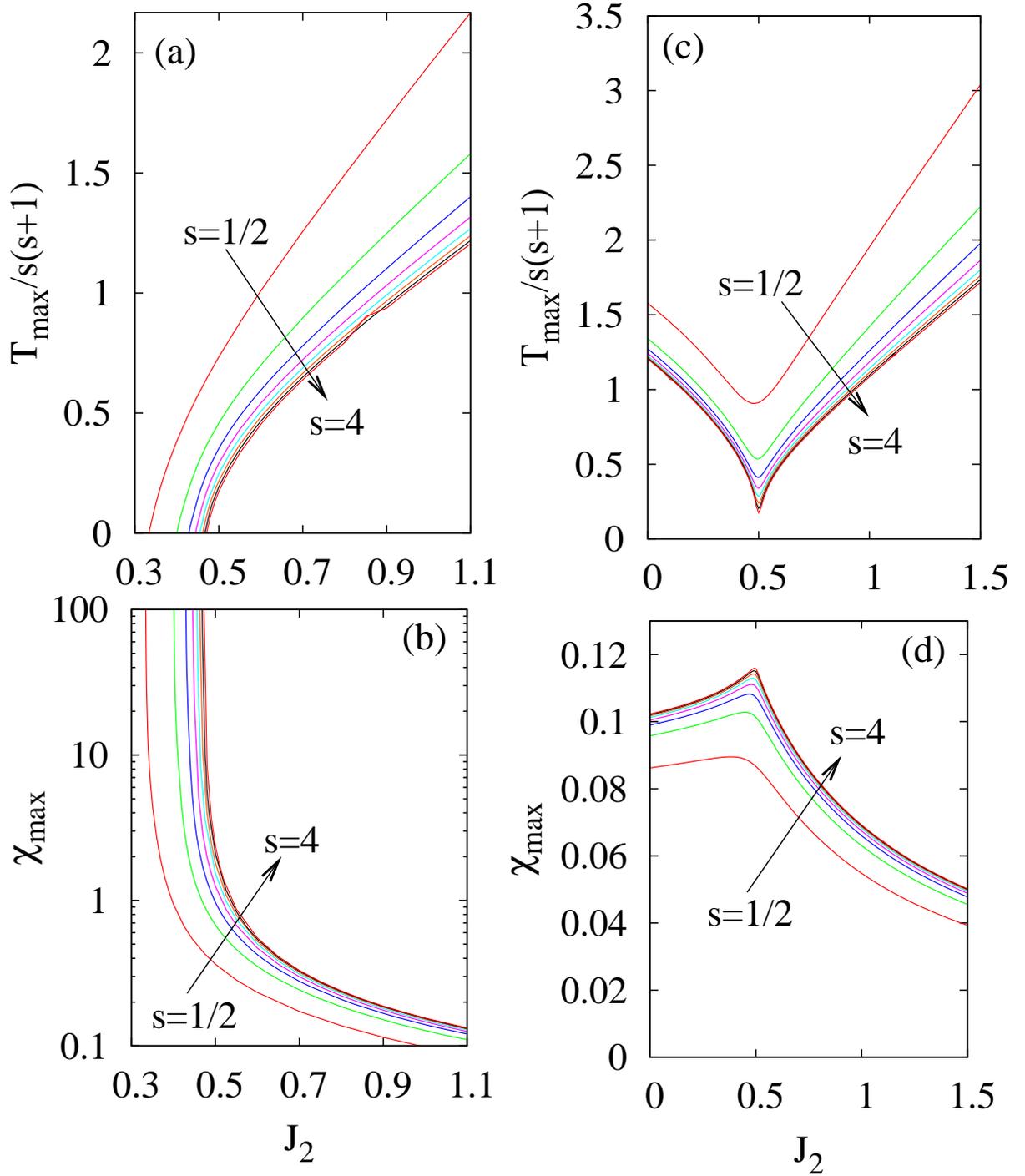}
\caption{\label{fig2}Position  $T_{max}$ (a and c) and height $\chi_{max}$ (b and d) of
$\chi(T)$ for the
finite $N=8$ square-lattice $J_1$-$J_2$ model (left panels FM $J_1=-1$, right
panels
AFM $J_1=+1$).}
\end{figure}

\begin{figure}[ht]
\includegraphics[width=40pc]{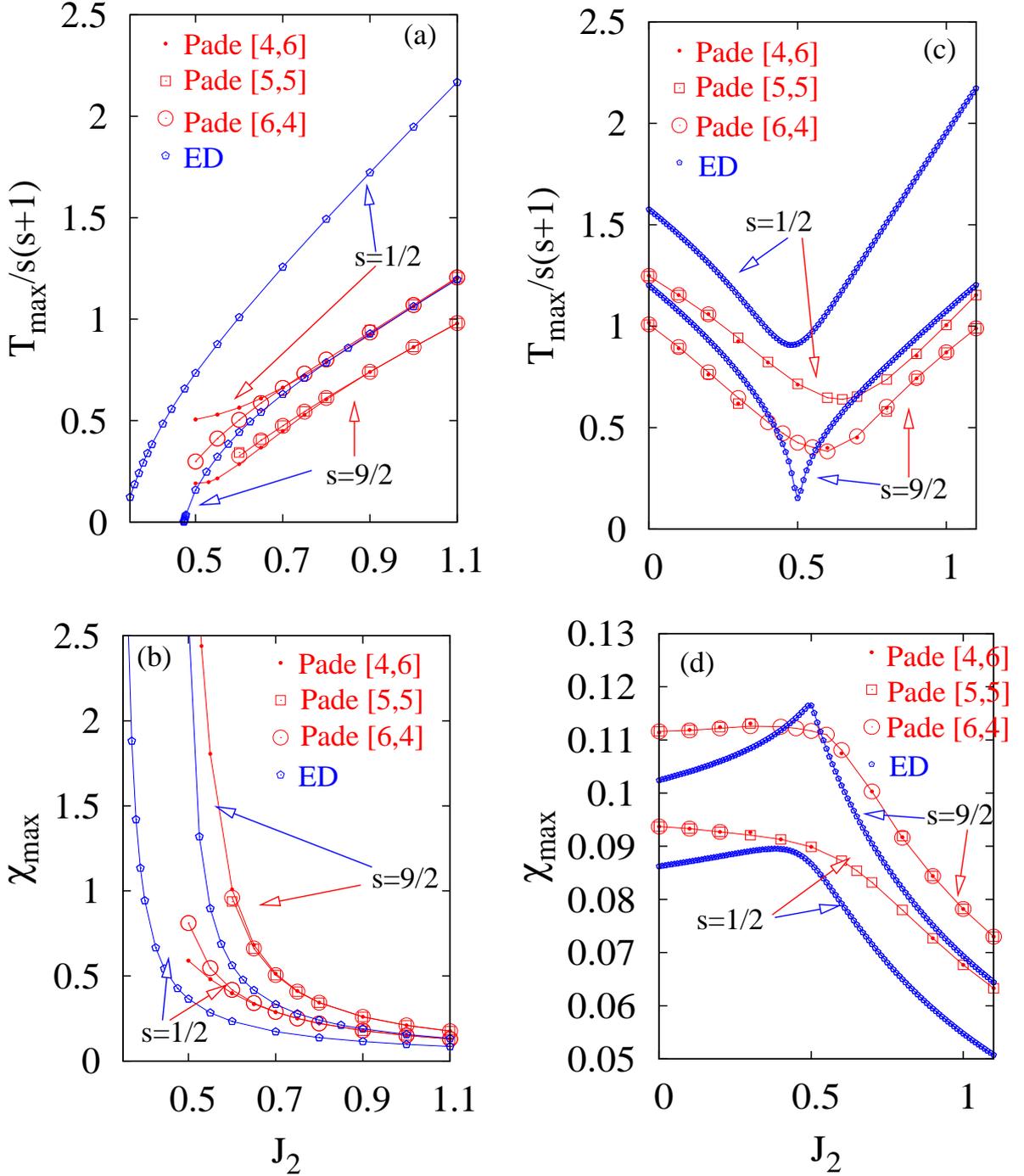}
\caption{\label{fig3}
Position  $T_{max}$ (a and c) and height $\chi_{max}$ (b and d)
 of
$\chi(T)$
for an infinite square-lattice $J_1$-$J_2$ model obtained by
10th  order HTE (left panels FM $J_1=-1$, right panels AFM $J_1=+1$). For
comparison
we show the ED data for $N=8$.}
\end{figure}

\begin{figure}[ht]
\includegraphics[width=38pc]{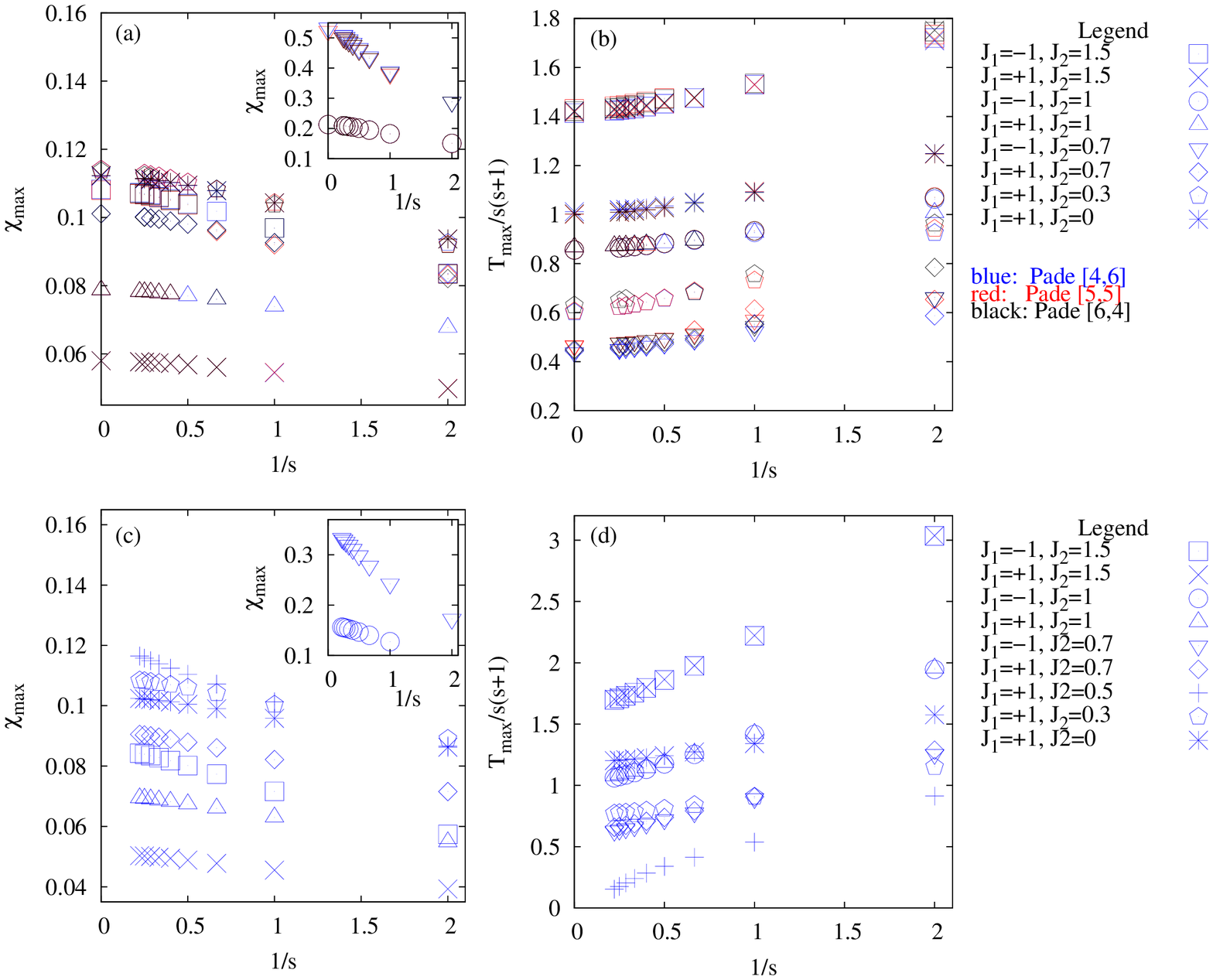}
\caption{\label{fig4}Height $\chi_{max}$ (a and c) and position  $T_{max}$
(b and d) of
$\chi(T)$ for the square-lattice $J_1$-$J_2$ model for various sets of parameters
$J_1$ and $J_2$ as a function of the inverse spin quantum number $s$ obtained
from Pad\'e approximants of 10th order HTE series of an infinite system (a and b) and from ED
for $N=8$ (c and d).}
\end{figure}

\section{Results}
First we present the temperature dependence of the
susceptibility $\chi$ in Fig.~\ref{fig1} for a particular value of $J_2$
and both FM and AFM $J_1$. In this paper the symbol $\chi$ means
$\chi|J_1|/Ng^2\mu_B^2$, where $N$ is the number of spins and $\mu_B$ is the
Bohr magneton.   The temperature is measured in terms of $|J_1|$, i.e. the
symbol $T$ means $T/|J_1|$.   
The qualitative behavior of $\chi(T)$ shown in Figs.~\ref{fig1}(a-d) is similar, there is the
broad maximum in $\chi(T)$ that is typical for a two-dimensional antiferromagnet (note
that for $J_2/|J_1|=1$ the system is
in the AFM ground state irrespective of the sign of $J_1$).
The  various $\chi(T)$ curves give an impression on  the finite-size effects, the
effect of the sign of the NN exchange $J_1$, and the influence of spin quantum
number $s$.
The height, $\chi_{max}$, and  the
position, $T_{max}$, of the maximum in the $\chi(T)$ curve
are interesting features
for the comparison with experimental data, in particular to get information
on the ratio $J_2/|J_1|$ from susceptibility measurements, see e.g.
\cite{13}.
Therefore we will discuss  $\chi_{max}$ and  the
$T_{max}$ now in more detail.

We present our data  for the susceptibility maximum for both  FM and
AFM NN exchange $J_1$ in Figs.~\ref{fig2} (ED data) and \ref{fig3}
(HTE and ED data). For FM $J_1$, $\chi_{max}$ ($T_{max}$) becomes
larger (smaller) upon lowering $J_2$. Finally, when approaching the
critical value $J_2^c$, where the transition to the ferromagnetic
ground state takes place, $\chi_{max}$ diverges  and $T_{max}$  goes
to zero. The critical point  for $s=1/2$ is $J^c_2=0.333\,|J_1|$ for
$N=8$ (but it is $J^c_2 \approx 0.4|J_1|$ for $N \to \infty$
\cite{12}). It increases with growing $s$ and becomes $J^c_2=0.5
|J_1|$ for $s \to \infty$. The data for $N=8$ and $N \to \infty$ are
in qualitative agreement. Although the finite-size effects are
obviously large, the general features of $\chi_{max}$  and $T_{max}$
as functions of $J_2$ and $s$ are quite similar. Naturally the HTE
fails when approaching $J^c_2$, since in this limit low temperatures
become relevant. Note that the HTE data for FM $J_1$ and $s=1/2$
are also in qualitative agreement with recently reported data
calculated by second-order Green's function approach \cite{13}. We
discuss now the case of AFM $J_1$ (right panels in Figs.~\ref{fig2}
and \ref{fig3}). For large $J_2$ the behavior of $\chi_{max}$ and
$T_{max}$ is very similar to that for FM $J_1$, i.~e.~the sign of
$J_1$ becomes irrelevant, cf.~Ref.~\cite{12}. On the other hand, for
smaller values of $J_2$ naturally both cases behave completely
different, since $J_1$ dominates the physics. We find a well
pronounced minimum in $T_{max}$ in the region of strongest
frustration around  $J_2=0.5$. For the finite system $\chi_{max}$
exhibits a maximum in this region, whereas  for the infinite system
$\chi_{max}$ is almost constant in the region $0 \le J_2 \le 0.5$.

To take a closer look on the role of the spin quantum number $s$ we present
in Fig.~\ref{fig4} the quantities $\chi_{max}$ and $T_{max}$ as a function of $1/s$ for
particular values of $J_2$.
Obviously, there is monotonous increase (decrease) of $\chi_{max}$
($T_{max}/s(s+1)$) with growing $s$.
For FM $J_1=-1$ the increase of $\chi_{max}$   is particular strong  for $J_2=0.7$ (see the
insets in panels a and c), since for large $s$ this value of $J_2$ becomes
quite close to the transition point to the FM ground state.
From Figs.~\ref{fig4}(a-d) it is also seen that the position $T_{max}$ of the maximum for
$J_2 \gtrsim  0.7|J_1|$ is almost independent of the sign of $J_1$, whereas
the height $\chi_{max}$ strongly depends on the sign of the NN coupling.
Let us finally mention the special $s$-dependence of the maximum in $\chi(T)$
for $J_1=1$ and $J_2=0.5$, where the classical ground state exhibits a
large non-trivial degeneracy.
The position $T_{max}/s(s+1)$ of the maximum shifts to zero in the limit $s
\to \infty$, whereas the height remains finite.
This behavior is quite similar to that found for the pyrochlore AFM
\cite{9,moessner99,huber01}, where the classical ground state is also highly
degenerate.

\section{Summary}\label{secIV}
Using high-temperature expansion and full exact diagonalization we have calculated the uniform
susceptibility
$\chi$ of
the spin-$s$ $J_1$-$J_2$ square-lattice Heisenberg magnet in a wide
parameter regime of FM and AFM $J_1$ and frustrating AFM $J_2$.
Especially, we have studied the height and the position of the
maximum in the $\chi(T)$ curve as functions of $J_2/J_1$ and the spin
quantum number $s$.
These data can be used
to get information on the ratio $J_2/|J_1|$ from susceptibility
measurements, e.g. on
oxovanadates which are well described by the square-lattice
$J_1$-$J_2$ model.

\ack
The work at Ames Laboratory was supported by the U.S. Department of Energy
under Contract No. DE-AC02-07CH11358.

\section*{References}
%%%%%%%%%%%%%%%%%%%%%%%%%%%%%%%%%%%%%%%%%%%

\end{document}